\documentclass[sigconf]{acmart}
\makeatletter
\renewcommand\@formatdoi[1]{\ignorespaces}
\makeatother
\usepackage{booktabs} 
\usepackage{color}
\usepackage{etoolbox}
\usepackage{indentfirst}

\setcopyright{rightsretained} 
\acmDOI{\url{https://doi.org/10.1145/3139958.3139968}}
\acmISBN{978-1-4503-5490-5/17/11}
\acmPrice{}
\copyrightyear{2017}
\acmConference[SIGSPATIAL'17]{ACM SIGSPATIAL}{November 7--10, 2017}{Los Angeles Area, CA, USA}

\makeatletter
\patchcmd{\maketitle}{\@copyrightspace}{}{}{}
\makeatother

\begin{document}
\pagenumbering{gobble} 
\title{GUIDES -- Geospatial Urban Infrastructure Data Engineering Solutions}


\author{Booma Sowkarthiga Balasubramani\textsuperscript{1}, Omar Belingheri\textsuperscript{1,2}, Eric S. Boria\textsuperscript{1}, Isabel F. Cruz\textsuperscript{1},\\Sybil Derrible\textsuperscript{1}, Michael D. Siciliano\textsuperscript{1}}
\affiliation{%
  \institution{\textsuperscript{1}University of Illinois at Chicago, USA}
}
\affiliation{\institution{\textsuperscript{2}University of Milan Bicocca, Italy}}
\email{bbalas3@uic.edu, o.belingheri@campus.unimib.it, {eboria2,ifcruz,derrible,siciliano}@uic.edu}






\begin{abstract}
\noindent
As the underground infrastructure systems of cities age, maintenance and repair become an increasing concern. Cities face difficulties in planning maintenance, predicting and responding to infrastructure related issues, and in realizing their vision to be a smart city due to their incomplete understanding of the existing state of the infrastructure. Only few cities have accurate and complete digital information on their underground infrastructure (e.g., electricity, water, natural gas) systems, which poses problems to those planning and performing construction projects. To address these issues, we introduce \emph{GUIDES} as a new data conversion and management framework for urban underground infrastructure systems that enable city administrators, workers, and contractors along with the general public and other users to query digitized and integrated data to make smarter decisions. This demo paper presents the \emph{GUIDES} architecture and describes two of its central components: (i) mapping of underground infrastructure systems, and (ii) integration of heterogeneous geospatial data.
\vspace{-7pt}
\end{abstract}
%

\vspace{4pt}
\maketitle
\thispagestyle{empty}
\vspace{-13pt}
\section{Introduction and Motivation} \label{introduction}
Complex geo-analytical applications require the integration of multiple cross-domain geospatial datasets, such as soil types, underground water pipes, and traffic conditions, which change with respect to time and space for effective spatial decision-making (e.g., identification of the most effective sites where to repair a water leakage). 
\emph{Geospatial data integration} involves combining two or more geospatial datasets from different sources to facilitate analysis, reasoning, querying, and data visualization.
\\\indent Significant opportunities for smarter data management of urban infrastructure systems are on the rise, as many US cities are moving towards the vision of ``smart cities'', 
creating open data portals that enable city administrators and residents to explore urban data and perform predictive analyses. Despite the availability of a tremendous volume of available data on cities, the lack of accurate geospatial data for underground infrastructure systems remains a problem. The need to address the poor state of the existing underground infrastructure is a strong rationale to develop such data management systems. For example, New York City has over 6,800 miles of water mains whose average age is 69 years. Over two thirds of them are made of materials susceptible to internal corrosion and prone to leakage, leading to 400 water main breaks in 2013 alone. Cross-domain querying is vital for an effective infrastructure maintenance (e.g., to locate pipes that need to be replaced in order of priority while  
coordinating across agencies to perform road excavation at the same time), 
and reiterate the need for integrating multiple heterogeneous geospatial datasets, thereby facilitating  queries such as \emph{retrieve all the components from the multiple thematic layers (e.g., census, water pipes, road network
)} in a given region, and \emph{how many low-income families will be affected by the burst of a given water main}. 
Such queries are complex to process due to various kinds of 
heterogeneities associated with them. 
Therefore, traditional ontology matching techniques, and the statistical and geospatial data processing tools (e.g., QGIS, ArcGIS) are insufficient to handle such queries.
\\\indent Data come from various sources, they possess differences in format, representation, context, tools, traits, structure, events, data models, spatio-temporal resolution, data collection and storage techniques, and the relationship between various system properties in a given region. Also, data are most often erroneous, incomplete, and inconsistent, leading to uncertainty. All these factors affect a data conversion and management framework, resulting in imprecise results when the data are analyzed.
\\\indent \emph{GUIDES} aims to enable a wide variety of users to explore and query underground infrastructure systems and analyze the impacts of disruptions in these systems (e.g., traffic conditions due to a water main break, malaria incidence in a county due to wastewater leakage), while addressing several technical challenges associated with achieving this vision, and protecting sensitive data simultaneously. 
In this paper, we describe \emph{GUIDES}, a novel framework to map and query urban underground infrastructure systems. We intend to demo the mapping, pre-processing, and part of the integration process  of \emph{GUIDES}, using the infrastructure data of the University of Illinois at Chicago (UIC) campus. 
\vspace{-9pt}
\section{Framework} \label{framework_new}
\begin{figure}[h]
\vspace{-8pt}
\includegraphics[width=3.3 in, height=2.1 in]{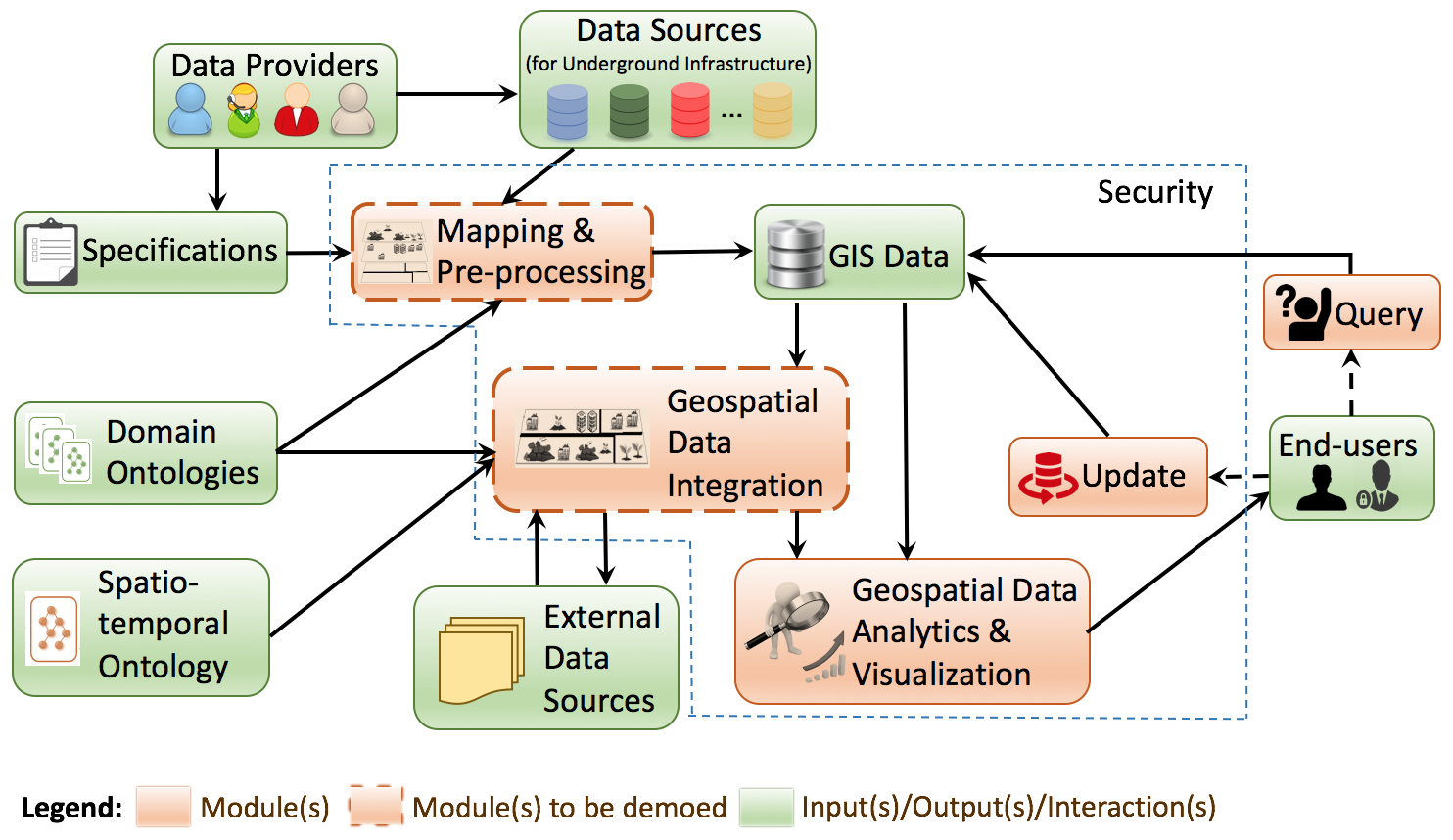}
\vspace{-8pt}
\caption{The GUIDES framework.}
\vspace{-15pt}
\label{fig:framework}
\end{figure}

This section introduces the \emph{GUIDES} framework (Figure~\ref{fig:framework}) and 
briefly describes its components.

\vspace{-10pt}
\subsection{Mapping} \label{preprocessing}
\vspace{-9pt}
\noindent \subsubsection{Data Sources and Data Providers}
Big-data driven decision making in smart city applications requires the integration of diverse map-based data sources, many of which are non-standardized. Standardization of data sources and coordination among \emph{data providers}, such as municipalities and service providers can improve the accuracy of data that is being centrally integrated. While some municipalities are working to verify map accuracy through on-field inspection and real-time sensor information, the accuracy of such geospatial data still remains problematic. 
An initial challenge of the \emph{GUIDES} framework is to create accurate GIS-based representations from existing legacy sources to enable the mapping of multiple thematic layers (e.g., buildings layer and water pipes layer).
\vspace{-6pt}
\noindent \subsubsection{Mapping \& Pre-processing}
Mapping deals with the conversion of data from one or more non-standardized sources into a single standardized format. Legacy data formats lack geographical information and often contain all the relevant information in one single source. Dimensions, for example, are often shown directly on the engineering drawing (e.g., CAD) as opposed to being an attribute of a piece of the infrastructure. Pre-processing algorithms that can automatically detect and solve these issues are critical. \emph{GUIDES} follows a three-step approach for pre-processing. First, 
a set of rules is developed based on domain knowledge to identify errors (e.g., two overlapping or co-located points should be flagged as they may be a single point). 
The second step is to generate new variables (e.g., based on network properties) that can be used to further identify errors (e.g., a water valve should have at least two connections). At the same time, we also incorporate GIS features to test whether a point is located within a polygon or not. After highlighting misplaced or missing elements, the third step is to suggest the correct configuration, for which, we develop algorithms and leverage the information present in other infrastructure systems. For instance, given that most underground infrastructure systems are buried under roads, road data can be used to suggest where missing infrastructure should be located. 
Once complete, all errors and added infrastructure elements can be flagged until they are validated manually during maintenance or new construction.
\vspace{-9pt}
\subsection{Geospatial Data Integration} \label{introduction}
Data may be collected with different spatial and temporal resolutions, update frequencies, and geometry types~\cite{Ref4} with heterogeneity across dimension, location, scale and source. To address these challenges, \emph{GUIDES} uses two kinds of ontologies: (i) a set of domain ontologies; and (ii) a spatio-temporal ontology. The domain ontology deals with instances related to a specific domain (e.g., water pipes) in the GIS database or relevant external data sources (e.g., census or economic data), whereas the spatio-temporal ontology consists only of the spatial (e.g., urban spatial hierarchy) and temporal (e.g., aggregation of monthly series to annual levels) hierarchies and their corresponding instances. Data integration is then carried out by performing \emph{instance matching}, which enables combining the datasets based on the similarity between their spatio-temporal components by matching their corresponding domain ontology with the spatio-temporal ontology.
\vspace{-10pt}
\subsection{Geospatial Data Analytics \& Visualization} \label{analytics}
The \emph{analytics} module incorporates geostatistical models and spatio-temporal processing mechanisms which  enable precise predictions of values for geospatial entities, and quantification of uncertainty. For example, this module applies the spatial function \emph{contains} to identify whether a census block contains a broken water pipe when computing the number of low-income families affected by a water main break in a given region. The \emph{visualization} module consists of a map-based interface for data exploration and comparison of various geostatistical models. Infrastructure elements can be displayed simultaneously for a given spatial entity (e.g., a street with several infrastructure elements including water pipes and buildings) to facilitate better decision making and data exploration with focus (e.g., details on an area where a water leakage is being repaired) and context (e.g., a sketch of other infrastructure elements around the focus area) at the same time.
\vspace{-10pt}
\subsection{Query \& Update} \label{query}
The \emph{query} module allows a wide range of geospatial queries for any spatial entity (e.g., census block, street, or a drawn extent) selected by users, whereas the \emph{update} module enables users to add, remove or modify the infrastructure elements in a dataset. Both the \emph{query} and \emph{update} modules restrict their allowed operations, depending on the category of the end-users (e.g., administrators, residents, maintenance crews) and their particular data needs and authorized level of access. For example, the general public should not be aware of the underground infrastructure data that are deemed sensitive, and hence are denied access to those data.
\vspace{-7pt}
\section{Demonstration Scenarios} \label{demo}
This section demonstrates how the \emph{GUIDES} framework enables pre-processing and ontology-based data integration mechanisms for urban infrastructure data, using the \emph{Water Pipes} and \emph{Buildings} maps of the UIC campus. 
These maps were initially in DWG (AutoCAD\footnote{\url{https://www.autodesk.com/products/autocad/}} drawing format), but were converted to shapefile format, and visualized using QGIS.\footnote{\url{http://qgis.org/}} The original data contained several errors and inconsistencies, and the conversion process generated several errors as well. The maps are transformed into a list of nodes and edges using a Python script~\cite{karduni2016protocol}, splitting the edges at the intersections with nodes.
\vspace{-8pt}
\subsection{Water Pipes Map Pre-processing}
\begin{figure}
\includegraphics[width=3in]{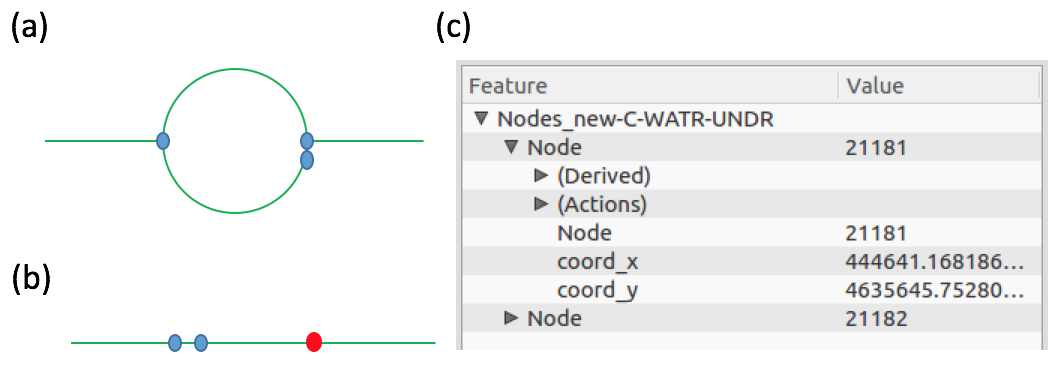}
\vspace{-10pt}
\caption{Errors in water pipes map.}
\label{fig:demo12}
\vspace{-10pt}
\end{figure}
This subsection details how \emph{GUIDES} facilitates identification and correction of errors in geospatial datasets.
\vspace{-6pt} 
\noindent \subsubsection{Fixing Duplicate Nodes}
In Figure~\ref{fig:demo12}b, although the feature highlighted in red appears to be a single node, the corresponding feature table in Figure~\ref{fig:demo12}c shows that it is in fact two nodes with two different IDs. That is, the two nodes are separate features within the same layer and there is no edge connecting them. Such scenarios are common and pose obvious issues, even when the most basic operations on the network are performed. For example, trying to find a path that goes through the edges in Figure~\ref{fig:demo12}b will fail, simply because the overlapping nodes are not connected. To resolve this, a Python script involving the \emph{GDAL}\footnote{http://www.gdal.org/} and \emph{NetworkX}\footnote{https://github.com/networkx/networkx} libraries is run to remove one node for each pair of such duplicate nodes and connect its edges to the other copy of the node.
\vspace{-5pt}
\subsubsection{Differentiating Infrastructure Elements and AutoCAD Symbols}
Figure~\ref{fig:demo12}a 
is an example of a circle representing a manhole. By zooming in, we can see that only one of the three nodes is actually connected to the circle edge. The circle was deleted and replaced with a new node at its center, with proper connections to the other nodes on either side. A field named \emph{Is\textunderscore manhole} with value $1$ for this node, is added in the attribute table of the map, so that the information is kept intact, even though the circle is removed.
\vspace{-17pt}
\noindent \subsubsection{Context-aware Pre-processing}
\begin{figure}[h]
\includegraphics[width=2.5in]{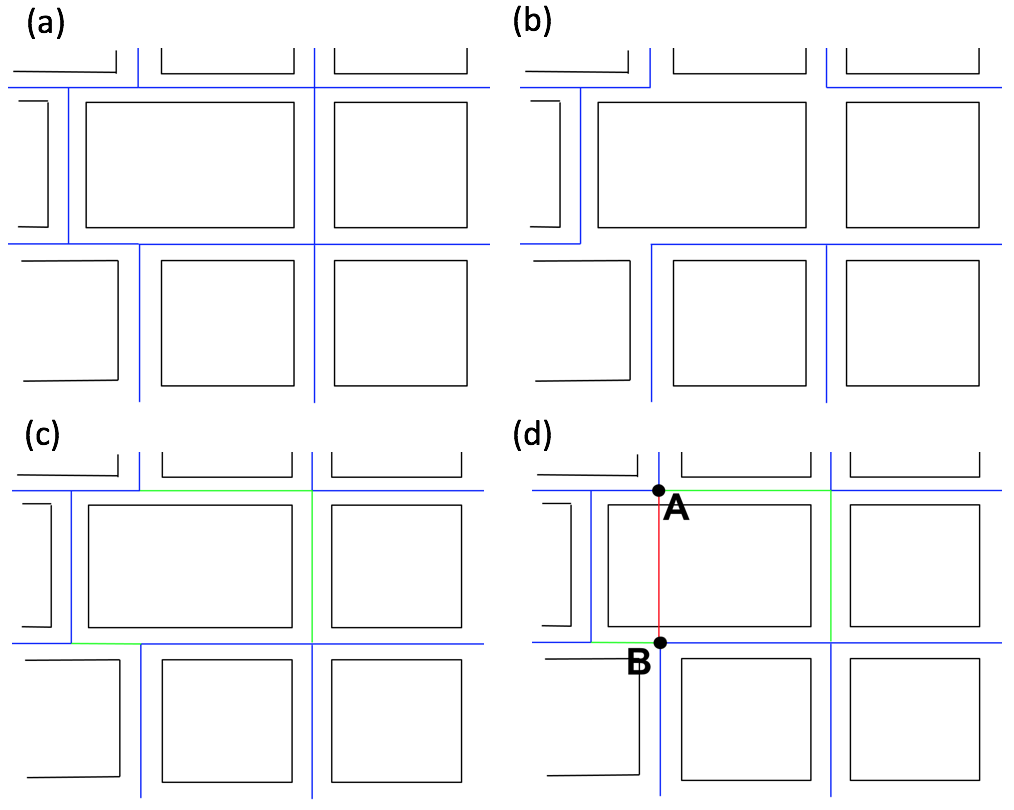}
\vspace{-10pt}
\caption{(a) Streets (black lines) and water pipes (blue lines); (b) After the removal of some pipes; (c) Missing pipes inferred by the algorithm; (d) A false positive (edge AB) was avoided, thanks to the streets layer.}
\label{fig:demo14}
\vspace{-11pt}
\end{figure}
The \emph{Buildings} layer was used to further identify errors in the \emph{Water Pipes} map. For example, intuitively, a water pipe should either end in a building, or be connected to other water pipes. Otherwise, 
it is reasonable to assume that there is an error that should be flagged for correction. Such cases can be identified by finding the nodes with degree $1$ (end nodes) in the \emph{Water Pipes} layer. 
This hypothesis has been confirmed by our experiments with synthetic map layers for \emph{Water Pipes} and \emph{Streets} (Figure~\ref{fig:demo14}). After the random removal of water pipes (Figure~\ref{fig:demo14}b), the algorithm suggested proper corrections to restore the initial map (Figure~\ref{fig:demo14}c). In doing so, using the constraints enforced by the \emph{Streets} layer (water pipes normally run underneath streets) has proven to be fundamental in reducing the number of false positives (incorrectly added pipes), raising the precision from 59\% to 93\%. Figure~\ref{fig:demo14}d shows an example of a pipe whose incorrect addition has been avoided with the help of these constraints.
\\\indent Applying this hypothesis to the UIC datasets, we should also ensure that the end nodes within the perimeter of a building are not flagged, which is essentially a \emph{point-in-polygon} problem~\cite{Sharma2014MethodsTD}. 
To resolve this, we use the \emph{GDAL} Python library, which, given a point (a node in the water pipes) and a polygon (a building), checks whether the the point falls within the area of the polygon.
\\\indent The \emph{GDAL} library allows for the creation of multipolygons, which are objects that can contain several polygons. With this feature, one object contained the polygons of all the buildings, instead of having one object (a polygon) for each building. The point-in-polygon check was then performed with the multipolygon in one iteration over the nodes, instead of using two iterations to check if any of the nodes (1st iteration) are in any of the buildings/polygons (2nd iteration). 
The solution with two iterations results in a much faster computation compared to the one with single iteration, and was therefore chosen for the final implementation.
\begin{figure}[h]
\vspace{-11pt}
\includegraphics[width=2.75in]{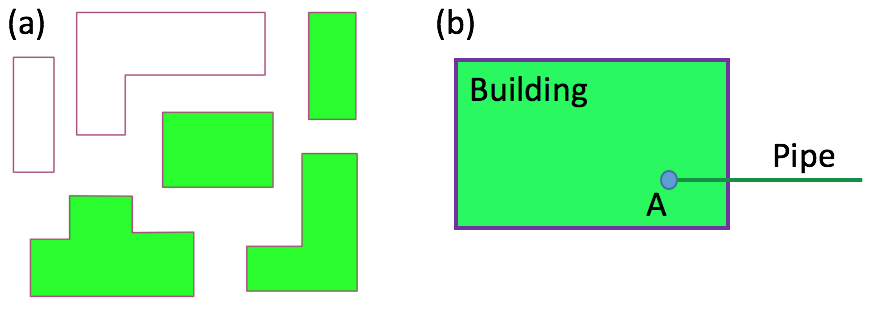}
\vspace{-11pt}
\caption{(a) Buildings (purple lines) \& corresponding polygons (green areas); (b) Sample water pipes \& buildings layer.}
\vspace{-14pt}
\label{fig:demo8}
\end{figure}
\\\indent From Figure~\ref{fig:demo8}a, we can see that the polygons (green areas) do not cover all of the buildings (purple lines) that the map contains because of the map inconsistencies (e.g., broken edges and detached nodes), which make it impossible for \emph{GUIDES} to build all the polygons properly. Therefore, this layer needs to be pre-processed to remove impurities and connect nodes that define the boundaries of a building, which we do by testing whether a \emph{building} node is on the edge of a full polygon or not, and if not, it is connected to the closest node and flagged. Figure~\ref{fig:demo8}b shows a water pipe entering a building. Although node $A$ has degree $1$, it will not be flagged as it lies within the area of the building. Implementation of machine learning algorithms (e.g., SVM) to identify inconsistencies in the data and suggest correct configurations is underway.
\vspace{-6pt}
\subsection{Ontologies in Geospatial Data Integration}
\begin{figure}[h]
\vspace{-9pt}
\includegraphics[width=2.5in, height = 1.25in]{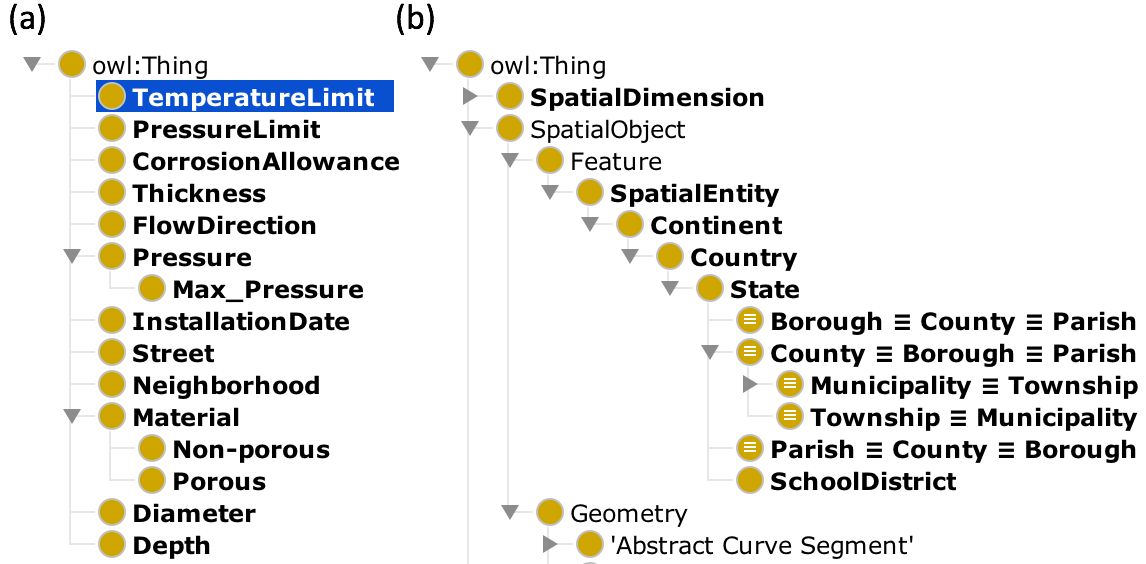}
\vspace{-10pt}
\caption{(a) Domain ontology (e.g., water pipes) - Partial; \\(b) Spatio-temporal ontology - Partial.}
\label{fig:demo11}
\vspace{-11pt}
\end{figure}
The integration component makes use of the pre-processed data and aids in resolving queries on location- and time-specific data. For example, given a region: (a) retrieve water pipes and buildings information; (b) retrieve all the components of the multi-layer network (road network, water pipes, rail network, and so on). Queries such as (a) facilitates the identification of potential spots where a water pipe is to be laid, when a new building is constructed. These queries are also particularly difficult to process, because of the heterogeneity of the spatial regions associated with the different datasets. For example, different infrastructure systems may belong to different spatial entities. Similarly, temporal queries also require the matching of heterogeneous data, mostly due to different temporal resolutions and update frequencies.
\begin{figure}[h]
\vspace{-8pt}
\includegraphics[width=2.65in]{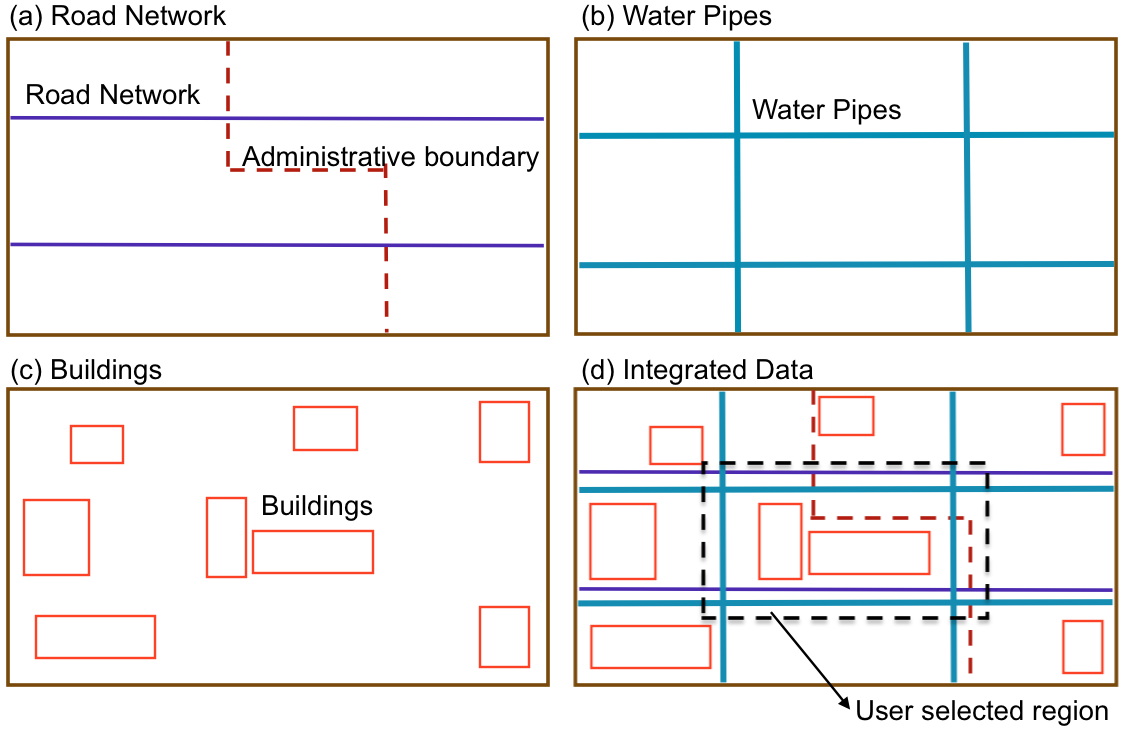}
\vspace{-10pt}
\caption{Geospatial data integration - An example.}
\label{fig:demo13}
\vspace{-10pt}
\end{figure}
\\\indent To retrieve all components of a multilayer network for a given region (Figure~\ref{fig:demo13}), we need to consider the different spatial and temporal resolutions they exhibit (e.g., road networks running across different cities, and water pipes managed individually by each city). To resolve such queries, we perform ontology-based geospatial data integration. \emph{GUIDES} encompasses a domain ontology for each dataset (e.g., water pipes as in Figure~\ref{fig:demo11}a), and a generic spatio-temporal ontology (Figure~\ref{fig:demo11}b), constructed using Prot\'{e}g\'{e}.\footnote{\url{http://protege.stanford.edu/}} Once the query is issued, the spatial and temporal components for the query are identified and their corresponding super- and sub-classes in the spatio-temporal ontology are obtained. We then retrieve the mappings (consisting of these super- and sub-classes) already obtained by matching the instances in the corresponding domain ontology 
with the instances of the spatio-temporal ontology, based only on the spatial and temporal components, using the AgreementMakerLight (AML)~\cite{faria2013agreementmakerlight} framework. Spatio-temporal functions such as \emph{within}, \emph{crosses}, 
are used to obtain query results only for the region and time selected.
\vspace{-8pt}
\section{Related Work} \label{relatedwork}
\vspace{-2pt}
GIVA~\cite{cruz2013giva}, an interactive map-based application, facilitates integration of data from multiple datasets, for a given region and a time interval.
\emph{GUIDES} adds on to the capabilities of GIVA, in terms of mapping and context-aware pre-processing, use of external data sources, and the mechanism for data integration, focusing on the urban and underground infrastructure domains. 
The City of Chicago's OpenGrid~\cite{Ref32}, a map-based open-source platform, supports advanced queries to identify and monitor incidents across the city. 
Howeer, it only accepts queries on limited datasets and 
does not support data integration, nor cross-domain querying, but can be extended to perform predictive analytics on urban data~\cite{balasubramani2016ontology}. 
SocialGlass~\cite{psyllidis2015platform} is a web-based system for visual exploration of large-scale and heterogeneous urban data, but it focuses on events and not the urban infrastructure.
Chang et al.~\cite{chang2007legible} propose a model for visualization of urban relationships using data aggregation techniques. Their model does not support geospatial data integration. 
The framework of Beck et al.~\cite{beck2007framework} integrates utility data using lightweight ontologies, but requires major changes when a new dataset needs to be integrated. In conclusion, real-life scenarios are more complex than previous work can handle, thus reinforcing the need for a new framework like \emph{GUIDES}.
\vspace{-8pt}
\section{Conclusions and Future Work} \label{conclusion}
\vspace{-2pt}
In this paper, we introduced \emph{GUIDES}, a data conversion and management framework, which supports ontology-based data integration, querying, analytics, and visualization of heterogeneous geospatial datasets, focusing on the urban infrastructure domain. The framework also supports several types of users such as administrator, planner, maintenance crew, and the general public, with various levels of access. We highlighted the key architectural elements and their capabilities to handle several challenges associated with geospatial data. Given the novelty of \emph{GUIDES} and the complexity of the problems this framework handles, there is a great potential for its expansion to ensure the highest level of usability and interoperability, cross-jurisdictional and inter-organizational collaboration, and workflow optimizations for crews.
Opportunities for integration of the \emph{GUIDES} framework with open data exploration platforms such as OpenGrid, will also be explored.
\vspace{-7pt}
\begin{acks}
\vspace{-2pt}
\noindent
We thank Roberto Tamassia and Goce Trajcevski for many helpful discussions. This work was partially supported by NSF awards CNS-1646395, III-1618126, CCF-1331800, and III-1213013 and by a Bill \& Melinda Gates Foundation Grand Challenges Explorations grant.
\end{acks}

\vspace{-6pt}
\bibliographystyle{acm}
\bibliography{sigspatial_main} 

\end{document}